\let\accentvec\vec
\documentclass[a4paper]{llncs}

\let\vec\accentvec

\usepackage[utf8]{inputenc}

\usepackage{amsmath,amssymb}

\usepackage{tikz}
\usetikzlibrary{backgrounds,automata,arrows,patterns,calc,positioning}

\usepackage[linesnumbered]{algorithm2e}

\usepackage{rotating}
\usepackage{hyperref}

\usepackage{picins}

\newcommand{\defo}[1]{{\color{blue}\emph{#1}}}

\newcommand{\pseudo}[1]{\texttt{#1}}

\newcommand{\mL}{\mathcal{L}}

\newcommand{\st}{{s\!t}}
\newcommand{\prg}{\mathcal{P}}
\newcommand{\program}{\prg}
\newcommand{\ellinit}{\ell_\mathsf{init}}
\newcommand{\mI}{\mathcal{I}}

\newcommand{\stsm}[1]{\text{\small \tikz[baseline]{\node[shape=rectangle,rounded corners=1.5mm,draw=lightgray,fill=lightgray!90,inner sep=0pt] at (0,.64ex){\hspace{.3em}\texttt{\upshape\strut#1}\hspace{.3em}\strut};}}}
\newcommand{\stsc}[1]{\tikz[baseline]{\scriptsize\node[shape=rectangle,rounded corners=1.5mm,draw=lightgray,fill=lightgray!90,inner sep=0pt] at (0,.64ex){\hspace{.3em}\texttt{\strut#1}\hspace{.3em}\strut};}}
\newcommand{\stsccol}[2]{\tikz[baseline]{\scriptsize\node[shape=rectangle,rounded corners=1.5mm,draw=#2,fill=#2,inner sep=0pt] at (0,.64ex){\hspace{.3em}\texttt{\strut#1}\hspace{.3em}\strut};}}
\newcommand{\var}{\mathsf{v}}
\newcommand{\oldrk}{\texttt{oldrnk}}
\newcommand{\valoldrk}{oldrnk}
\newcommand{\rankDecrease}{{\texttt{rankDecrease$_{uvf}$}}}

\newcommand{\WF}{\mathbb{W}}

\newcommand{\Psort}{\mathcal{P}^\texttt{sort}}

\newcommand{\vardom}{\mathsf{dom}}
\newcommand{\Loc}{\mathsf{Loc}}
\newcommand{\ellfin}{\ell_\mathsf{fin}}

\newcommand{\hoare}[3]{ \{ \; #1 \; \} \; #2 \; \{ \; #3 \; \}}

\definecolor{inner}{RGB}{180,180,220}%alternative 19,0,128
\definecolor{innerT}{RGB}{00,00,120}
\definecolor{outer}{RGB}{180,220,180}%alternative 83,128,0
\definecolor{outerT}{RGB}{00,120,00}

\newcommand{\Inner}{\textsc{\color{innerT}Inner}}
\newcommand{\Outer}{\textsc{\color{outerT}Outer}}

\begin{document}

\title{Termination Analysis\\ by Learning Terminating Programs
\thanks{The final publication is available at \url{http://link.springer.com}.}
\thanks{This work is supported by the
  German Research Council (DFG) as part of the Transregional Collaborative
  Research Center ``Automatic Verification and Analysis of Complex Systems''
  (SFB/TR14 AVACS)}
}

\author{Matthias Heizmann\and Jochen Hoenicke \and Andreas Podelski}

\institute{University of Freiburg, Germany}

\maketitle

% Topics:
%
%  Abstract interpretation, Model checking, Modularity, Static analysis,  Temporal properties/Termination, Verification
%  Synthesis/Generation

\begin{abstract}
We present a novel approach to termination analysis. In a first step, the analysis uses a program as a black-box which exhibits only a finite set of sample traces. Each sample trace is infinite but can be represented by a finite lasso. The analysis can "learn" a program from a termination proof for the lasso, a program that is terminating by construction. In a second step, the analysis checks that the set of sample traces is representative in a sense that we can make formal. An experimental evaluation indicates that the approach is a potentially useful addition to the portfolio of existing approaches to termination analysis. 
\end{abstract}

\section{Introduction}
 
Termination analysis is an active research topic, and a wide range of methods and tools exist~\cite{pldi/CookPR06,conf/tacas/CookSZ13,sas/HarrisLNR10,cav/KroeningSTW10,cav/LeeWY12,tacas/PopeeaR12,conf/esop/UrbanM14}. Each method provides its own twist to address the same issue: in the presence of % \emph{multi-path} loops (i.e.,
loops with branching or nesting, the termination argument has to account for all possible interleavings between the different paths through the loop. 

If the program is \emph{lasso-shaped} (a stem followed by a single loop without branching), the control flow is trivial; there is only one path.
Consequently, the termination argument can be very simple.  Many procedures are specialized to lasso-shaped programs and derive a simple termination argument rather efficiently~\cite{cav/Ben-Amram09,popl/Ben-AmramG13,cav/BradleyMS05,tacas/CookKRW10,conf/atva/HeizmannHLP13,conf/tacas/LeikeH14,vmcai/PodelskiR04}. The relevance of lasso-shaped programs stems from their use as the representation of an infinite trace through the control flow graph of a program with arbitrary nesting.

% In this paper, we introduce the technique of \emph{control flow   refinement with fairness}.  The technique consists of transforming a general program (with arbitrary loops) $\program$ into a semantically equivalent program of a very specific form: $\mathtt{choose}(\program_1,\ldots,\program_n)$, the nondeterministic choice of finitely many programs, each of which comes with the same termination argument as a lasso-shaped program.

We present a new method that analyzes termination of a general program $\prg$ but has to find termination arguments only for lasso-shaped programs.
In our method we see the program $\prg$ as a blackbox from which we can obtain sample traces. 
We transform a sample trace $\pi_i$ into a lasso-shaped program and use existing methods to compute a termination argument for this lasso-shaped program.  
Afterwards we construct a ``larger'' program $\prg_i$ (which may have branching and nested loops) for which the same termination argument is applicable.
We call this construction \emph{learning}, because we learned the terminating program $\prg_i$ from a sample trace $\pi_i$.
Our algorithm continues iteratively until we learned a set of programs $\program_1,\ldots,\program_n$ that forms a decomposition of the original program $\prg$.
This decomposition can be seen as a program of the form $\mathtt{choose}(\program_1,\ldots,\program_n)$, i.e., a nondeterministic choice of programs $\program_1,\ldots,\program_n$, that is semantically equivalent to the original program $\prg$.

Our technical contribution is this method, which does not only extend the existing portfolio of termination analyses but also provides a new functionality: the decomposition of a program $\prg$ into modules $\prg_1,\ldots,\prg_n$.  
This decomposition is not guided by the syntax of the program, this decomposition exploits a novel notion of modularity where a module is defined by a certain termination argument. This novel notion of modularity is the conceptual contribution of our paper.
% Our approach extends the existing portfolio of termination approaches. Furthermore we provide an additional functionality which is the decomposition of a program into modules. Our algorithm provides this decomposition even if we are unable to prove termination or if the program is not terminating. In both cases we provide terminating modules $\program_1,\ldots,\program_{n-1}$ and one additional module $\program_n$ that is not terminating or for which termination is unknown.

\begin{center}
\begin{minipage}{40mm}
\pseudo{program sort(int i)}
\begin{tabular}{@{}l@{\;}l@{}}
% $\ell_0$: & \pseudo{\large i:=n}\\[0.8mm]
$\ell_1$: & \pseudo{while (i>0)}\\[0.8mm]
$\ell_2$: & $\;\quad$\pseudo{int j:=1}\\[0.8mm]
$\ell_3$: & $\;\quad$\pseudo{while(j<i)}\\[0.8mm]
 & $\;\quad$\pseudo{//} $\quad$\pseudo{if (a[j]>a[i])}\\[0.8mm]
 & $\;\quad$\pseudo{//} $\quad\;\quad$\pseudo{swap(a[j],a[i])}\\[0.8mm]
$\ell_4$: & $\;\quad\;\quad$\pseudo{j++}\\[0.8mm]
$\ell_5$: & $\;\quad$\pseudo{i--}\\[0.8mm]
\end{tabular}
\end{minipage}\hspace{15mm}
\begin{minipage}{40mm}
\centering
\begin{tikzpicture}[scale=1.1,
zustd/.style={circle,draw,minimum size=10pt},
%trans/.style={>=stealth,thick}
trans/.style={thick,->}
]
\node (0) [zustd] at (0,0) {$\ell_1$};
\node (2) [zustd] at ($(0)+(1,-1)$) {$\ell_2$};
\node (3) [zustd] at ($(0)+(0,-2)$) {$\ell_3$};
\node (4) [zustd] at ($(0)+(0,-3.7)$) {$\ell_4$};
\node (5) [zustd] at ($(0)+(-1,-1)$) {$\ell_5$};

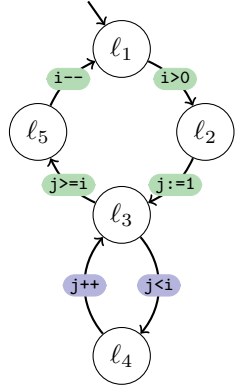
\draw [trans] ($(0) + (125:0.7)$) to (0);
\draw [trans,bend left=20] (0) to node {$\stsccol{i>0}{outer}$} (2);
\draw [trans,bend left=20] (2) to node {$\stsccol{j:=1}{outer}$} (3);
\draw [trans,bend left=40] (3) to node {$\stsccol{j<i}{inner}$} (4);
\draw [trans,bend left=40] (4) to node {$\stsccol{j++}{inner}$} (3);
\draw [trans,bend left=20] (3) to node {$\stsccol{j>=i}{outer}$} (5);
\draw [trans,bend left=20] (5) to node {$\stsccol{i--}{outer}$} (0);
\end{tikzpicture}
\end{minipage}
\end{center}
% \caption{Program $\prg^\texttt{sort}$ which is an implementation of bubblesort}
% \end{figure}

Let us explain our algorithm informally using the program $\prg^\texttt{sort}$ depicted above which is an implementation of bubblesort.
Termination of $\prg^\texttt{sort}$ can be shown, e.g.,  by using the quadratic ranking function
$f(\pseudo{i},\pseudo{j}) = \pseudo{i}^2-\pseudo{j},$
or the lexicographic ranking function
$f(\pseudo{i},\pseudo{j}) = (\pseudo{i},\pseudo{i}-\pseudo{j}).$
Intuitively, neither of the two ranking functions is a simple
termination argument.

Now, let us pick some $\omega$-trace from $\prg^\texttt{sort}$.  
We take the trace that first enters the outer while loop and then takes the inner while loop infinitely often.  
We denote this trace using the $\omega$-regular expression $\Outer. \Inner^\omega$.  
We see that this trace is terminating. Its termination can be shown using the linear ranking function $f(\pseudo{i},\pseudo{j})=\pseudo{i}-\pseudo{j}$.  
Moreover, we see that this ranking function is not only applicable to this trace, this ranking function is applicable to all traces that eventually always take the inner loop.
\begin{align}
(\textsc{\color{innerT}Inner}+\textsc{\color{outerT}Outer})^*. \textsc{\color{innerT}Inner}^\omega 
\end{align}

Now, let us pick another $\omega$-trace from $\prg^\texttt{sort}$.  
This time we take the trace that always takes the outer while loop. %, denoted by $\Outer^\omega$.  
We see that this trace is terminating. Its termination can be shown using the linear ranking function $f(\pseudo{i},\pseudo{j})=\pseudo{i}$.  
Moreover, we see that this ranking function is not only applicable to this trace, this ranking function is applicable to all traces that take the outer while loop infinitely often.
\begin{align}
(\Inner^*.\Outer)^\omega
\end{align}

Finally, we consider the set of all $\omega$-trace of the program $\Psort$
$$(\Outer + \Inner)^\omega,$$
check that each trace has the form (1) or has the form (2), and conclude that $\Psort$ is terminating.

If we are to automate the reasoning from 
the example above, a number of questions arise. 

\emph{(A) How does one effectively represent a set of traces that share a common reason for termination, like the sets (1) and (2) above?}
The answer is given in Section~\ref{sec:decomposition} where we define a \emph{module}, which is a program whose traces adhere to a certain fairness constraint.

\emph{(B) What is a termination argument whose applicability to a whole set of traces can be checked effectively?}
The answer is given in Section~\ref{sec:CertifiedModule} where we present a Floyd-Hoare style annotation for termination proofs.

\emph{(C) How can we learn a set of terminating traces (represented as a program with a fairness constraint) from a single terminating sample trace?}
The answer is given in Section~\ref{sec:automatic-module} where we construct a terminating module from a given termination proof.

\emph{(D) How can we check that a set of modules $\program_1,\ldots,\program_n$ covers the behavior of the original program $\prg$ and can we always decompose $\prg$ into a set of modules $\program_1,\ldots,\program_n$?}
One facet of the question is the theoretical completeness, which is answered in Section~\ref{sec:automatic-decomposition}. The other facet is the practical feasibility, which is analyzed via an experimental evaluation in Section~\ref{sec:evaluation}.

\section{Fair module}
\label{sec:decomposition}

\paragraph{Preliminaries.}
The key concept in our formal exposition is the notion of an
\defo{$\omega$-trace}, which is an infinite sequence of program
statements
$\pi=\st_1\st_2\ldots$.
We assume that the statements are taken from a given finite set of program statements $\Sigma$. 
If we consider $\Sigma$ as an alphabet and each statement as a letter,
then an \defo{$\omega$-trace} 
is an infinite word over this alphabet.
In order to stress the usage of statements as letters of an alphabet,
we sometimes frame each statement/letter.  For example, we can write
the alphabet of our running example $\Psort$ as
$
\Sigma_{sort}=\{ \stsccol{i>0}{outer}, \stsccol{j:=1}{outer}, \stsccol{j<i}{inner}, \stsccol{j++}{inner}, \stsccol{j>=i}{outer}, \stsccol{i--}{outer}\}
$
and
$
\pi = \stsccol{j<i}{inner} \stsccol{j:=1}{outer} . ( \stsccol{j:=1}{outer} \stsccol{j++}{inner} \stsccol{j:=1}{outer} )^\omega
$
is an $\omega$-trace. 

The definition of an $\omega$-trace as an arbitrary (infinite) sequence means
that the notion is independent of the programming language
semantics,  % In particular, a trace may not correspond to any possible execution.
which we even have not introduced yet.  We will do so now.
A \defo{valuation} $\nu$ is a function that maps the program variables $\vec\var$ to values.
We use the term \emph{valuation} instead of \emph{state} to stress that this is independent from the program counter (and independent from control flow).
We call a set of valuations a \defo{predicate} and use the letter $I$ to denote predicates.
The letter $I$ is used reminiscent to \emph{invariant}, because we will use predicates to represent invariants at locations.
We assume that each statement $\st$ comes with a binary relation over the set of valuations (the set of its precondition/postcondition pairs). 
We say that the Hoare triple $\{ I \} \st \{ I'\}$ is valid, if the binary relation for $\st$ holds between precondition $I$ and postcondition $I'$.
We use the interleaved sequences of valuations and statements
$\nu_0\stackrel{\st_{1}}{\rightarrow}\ldots\stackrel{\st_{n}}{\rightarrow}\nu_n$
as a shorthand to denote that each pair of valuations $(\nu_i,\nu_{i+1})$ is contained in the transition relation of the statement $\st_{i+1}$.

An $\omega$-trace may not correspond to any possible execution
for one out of two reasons. 
First, there may be a finite prefix that does not have any possible execution, like e.g., the prefix $\stsc{x<0}\stsc{x:=1}\stsc{x<0}$ of the $\omega$-trace $(\stsc{x<0} \stsc{x:=1})^\omega$. Secondly, there may be no starting valuation $\nu_0$ for any infinite execution, although every finite prefix is executable which holds e.g., for the $\omega$-trace $(\stsc{x>=0} \stsc{x--})^\omega$. In both cases we call such an $\omega$-trace \defo{terminating}.

The notion of an $\omega$-trace is also independent of a program (a trace may
not correspond to a path in the program's control flow graph).  
We introduce a program as a control flow graph whose edges are labeled with statements. Formally, a \defo{program} is a graph $\prg=\langle \Loc,\delta,\ellinit\rangle$ with a finite set $\Loc$ of nodes called \emph{locations}, a set $\delta$ of  edges labeled with statements, i.e.,
$\delta\subseteq \Loc\times\Sigma\times \Loc$ and an initial node called the initial location $\ellinit$. 
We call the program $\prg$ \defo{terminating} if each of its $\omega$-traces is terminating.

\paragraph{Module: program with fairness contraint.}
In our method we will decompose a program into \emph{modules} such that each module represents traces that share a common reason for termination. We now formalize our notion of a module.

\begin{definition}[module]
A \defo{module} is a program together with a fairness constraint
given by a distinguished
location  $\ellfin$, i.e., 
$$\prg=\langle \Loc,\delta,\ellinit,\ellfin\rangle$$
where the set of location can be partitioned into two disjoint sets, $\Loc_U$, and $\Loc_V$, such that
\begin{itemize}
\item the initial location is contained in $\Loc_U$, 
\item the final location is contained in $\Loc_V$, and
\item no location in $\Loc_V$ has a successor in $\Loc_U$, i.e.,
$$(\ell,\st,\ell')\in \delta \quad\text{\ \ implies \ \ }\quad \ell\in \Loc_U \text{ \  or \  } \ell'\in \Loc_V$$
\end{itemize}
A \defo{fair $\omega$-trace} of a module $\mathcal{P}$ is an
$\omega$-trace that labels a fair path in the graph of
$\mathcal{P}$, which is a path that visits the distinguished location
$\ellfin$ infinitely often.
We call the module $\prg$ \defo{terminating} if each of its fair $\omega$-traces is terminating. 
\end{definition}

A \emph{non-fair} $\omega$-trace of a terminating module (i.e., an
$\omega$-trace that labels a path in its control
flow graph without satisfying the fairness constraint) can be non-terminating.

For the reader who is familiar with the concept of Büchi automata, a
module is reminiscent of a Büchi automaton with exactly one final
state.  A Büchi automaton of this form recognizes an
$\omega$-regular language of the form $U.V^\omega$, where $U$
and $V$ are regular languages over the alphabet of statements
$U,V\subseteq\Sigma^*$.

\parpic[r]{
\scalebox{0.9999}{
\begin{tikzpicture}[scale=1.1,
zustd/.style={circle,draw,minimum size=10pt},
%trans/.style={>=stealth,thick}
trans/.style={thick,->}
]
\node (0) [zustd] at (0,0) {$\ell_0$};
\node (2) [zustd] at ($(0)+(1,-1)$) {$\ell_2$};
\node (3) [zustd] at ($(0)+(0,-2)$) {$\ell_3$};
\node (4) [zustd] at ($(0)+(0,-3.7)$) {$\ell_4$};
\node (5) [zustd] at ($(0)+(-1,-1)$) {$\ell_5$};
\node (3') [zustd,accepting] at ($(0)+(2,-2)$) {$\ell_3'$};
\node (4') [zustd] at ($(0)+(2,-3.7)$) {$\ell_4'$};
\draw [trans] ($(0) + (125:0.7)$) to (0);
\draw [trans,bend left=20] (0) to node {$\stsccol{i>0}{outer}$} (2);
\draw [trans,bend left=20] (2) to node {$\stsccol{j:=1}{outer}$} (3);
\draw [trans,bend left=40] (3) to node {$\stsccol{j<i}{inner}$} (4);
\draw [trans,bend left=40] (4) to node {$\stsccol{j++}{inner}$} (3);
\draw [trans,bend left=20] (3) to node {$\stsccol{j>=i}{outer}$} (5);
\draw [trans,bend left=20] (5) to node {$\stsccol{i--}{outer}$} (0);
\draw [trans] (2) to node {$\stsccol{j:=1}{outer}$} (3');
\draw [trans,bend left=40] (3') to node {$\stsccol{j<i}{inner}$} (4');
\draw [trans,bend left=40] (4') to node {$\stsccol{j++}{inner}$} (3');
\end{tikzpicture}
}
\scalebox{0.9999}{
\begin{tikzpicture}[scale=1.1,
zustd/.style={circle,draw,minimum size=10pt},
%trans/.style={>=stealth,thick}
trans/.style={thick,->}
]
\node (0) [zustd,accepting] at (0,0) {$\ell_0$};
\node (2) [zustd] at ($(0)+(1,-1)$) {$\ell_2$};
\node (3) [zustd] at ($(0)+(0,-2)$) {$\ell_3$};
\node (4) [zustd] at ($(0)+(0,-3.7)$) {$\ell_4$};
\node (5) [zustd] at ($(0)+(-1,-1)$) {$\ell_5$};
\draw [trans] ($(0) + (125:0.7)$) to (0);
\draw [trans,bend left=20] (0) to node {$\stsccol{i>0}{outer}$} (2);
\draw [trans,bend left=20] (2) to node {$\stsccol{j:=1}{outer}$} (3);
\draw [trans,bend left=40] (3) to node {$\stsccol{j<i}{inner}$} (4);
\draw [trans,bend left=40] (4) to node {$\stsccol{j++}{inner}$} (3);
\draw [trans,bend left=20] (3) to node {$\stsccol{j>=i}{outer}$} (5);
\draw [trans,bend left=20] (5) to node {$\stsccol{i--}{outer}$} (0);
\end{tikzpicture}
}
}
\begin{example}
Let us consider again our running example $\Psort$.
The sets that we gave informally by the $\omega$-regular expressions (1) and (2) can be represented as modules.
The module $\Psort_1$ depicted on the left represents all traces that eventually only take the inner while loop.
The module $\Psort_2$ depicted on the right represents all traces that take the outer while loop infinitely often.
\end{example}

In this example, the decomposition of the program into modules is defined by the nestings structure of while loops.
In Section~\ref{sec:automatic-decomposition} we present an algorithm that finds a decomposition automatically but does not rely on any information about the structure of the while loops in the program.

\section{Certified Module}
\label{sec:CertifiedModule}
In this section we present a termination argument for modules that consists of two parts: a ranking function and an annotation of the module's locations.

First, we extend the usual notion of a ranking function to our definition of a module. The crux in the following definition lies in the fact that we do not require that the value of the ranking function has to decrease after a fixed number of steps. We only require that the value of the ranking function has to decrease every time the final location $\ellfin$ is visited.
As a consequence our ranking function is a termination argument that is applicable to each fair $\omega$-trace, but does not have to take non-fair $\omega$-traces into account.

\begin{definition}[ranking function for a module]
Given a module $\prg$,
we call a function $f$ from valuations into a well-ordered set  $(\WF,\prec)$ a \defo{ranking function} for $\prg$ if for each 
finite path
$$\ell_0\stackrel{\st_{1}}{\rightarrow}\dots\stackrel{\st_{k}}{\rightarrow}\ell_k\stackrel{\st_{k+1}}{\rightarrow}\cdots\stackrel{\st_n}{\rightarrow}\ell_n$$
that starts in the initial location (i.e., $\ell_0=\ellinit$) and visits the final location in the $k$-th step and in the $n$-th step (i.e., $\ell_k=\ell_n=\ellfin$)
and for each sequence of valuations $\nu_0,\dots,\nu_n$ such that the pair $(\nu_i,\nu_{i+1})$ is in the transition relation of the statement $\st_i$, i.e.,
$$\nu_0\stackrel{\st_{1}}{\rightarrow}\dots\stackrel{\st_{k}}{\rightarrow}\nu_k\stackrel{\st_{k+1}}{\rightarrow}\cdots\stackrel{\st_n}{\rightarrow}\nu_n$$
the value of the ranking function decreases whenever $\ellfin$ is visited, i.e.,
$$f(\nu_n)\prec f(\nu_k).$$
\end{definition}

In all the following examples we take $\mathbb{Z}$ as domain of the program variables. Our well-ordered set $\WF$ will be $(\mathbb{Z}\cup\{\infty\},\prec)$. The ordering $\prec$ is the natural order restricted to pairs where the second operand is greater than or equal to zero (i.e., $a\prec b \;\text{ if and only if }\; a<b \;\land\; b\geq 0)$.

\begin{example}
The function $f:\vardom\to\mathbb{Z}\cup\{\infty\}$ defined as $f(i,j) = i-j$ is a ranking function for the module $\Psort_1$ depicted in Example~1.
\end{example}

\begin{lemma}\label{lem:rankingfunc}
If the module $\prg$ has a ranking function $f$, then each fair trace of the module is terminating. \footnote{An extended version of this paper that also contains the correctness proofs is available online.}
\end{lemma}

How can we check that a function is a ranking function for a module?
We next introduce a novel kind of annotation, called \emph{rank certificate} that serves as a proof for this task.
Informally, a \emph{rank certificate} is a Floyd-Hoare annotation that ensures that the value of the ranking function has decreased whenever the final location $\ellfin$ was visited. Therefore, we introduce an auxiliary variable $\oldrk$ that represents the value of the ranking function at the previous visit of $\ellfin$.
Initially, the auxiliary variable $\oldrk$ has the value $\infty$ which is a value strictly greater than all other values from our well-ordered $\WF$.

\begin{definition}[certified module] Given a module $\prg=\langle \Loc,\delta, \ellinit, \{\ellfin\} \rangle$
and a function $f$ from valuations into a well-ordered set $(\WF,\prec)$,
we call a mapping $\mI$ from locations to predicates
a \defo{rank certificate} for the function $f$ and the module $\prg$ if the following properties hold.

\begin{itemize}
  \item The initial location $\ellinit$ is mapped to the predicate where the auxiliary variable $\oldrk$ has the value $\infty$, i.\,e.,
 $$ \mI(\ellinit) \;\Leftrightarrow\; \oldrk = \infty.$$

 \item The accepting state is mapped to a predicate in which the value of the ranking function $f$ over the program variables is smaller than the value of the variable $\oldrk$, i.\,e., 
 $$ \mI(\ellfin) \;\Rightarrow\;  \big(f(\vec\var)\prec\oldrk\big).$$

 \item The outgoing edges of non-accepting locations correspond to valid Hoare triples, i.e.,
   $$\{ \; \mI(\ell) \; \} \; \st \;  \{ \; \mI(\ell') \; \} \text { is valid } 
\mbox{for }(\ell,\st,\ell')\in\delta, \ell\neq\ellfin
$$
 and outgoing edges of the final location correspond to valid Hoare triples if we insert an additional assignment statement that assigns the value of the ranking function to the auxiliary variable $\oldrk$ , i.e.,
   $$\{ \; \mI(\ell) \; \} \; \stsm{$\oldrk$:=$f(\vec\var)$};\st \;  \{ \; \mI(\ell') \; \} \text { is valid } 
\mbox{ for }(\ellfin,\st,\ell')\in\delta
$$
\end{itemize}
\noindent
We call the triple $(\prg,f,\mI)$ a \defo{certified module}.
\end{definition}

\parpic[r]{
\scalebox{0.95}{
\begin{tikzpicture}[scale=1,
zustd/.style={circle,draw,minimum size=10pt},
%trans/.style={>=stealth,thick}
trans/.style={thick,->}
]
\node (0) [zustd
,label=left:\text{\footnotesize $\begin{array}{c}\{\oldrk=\infty\}\end{array}$}
] at (0,0) {$\ell_1$};
\node (2) [zustd
,label=above right:\text{\footnotesize $\begin{array}{c}\{\oldrk=\infty\}\end{array}$}
] at ($(0)+(1,-1)$) {$\ell_2$};
\node (3) [zustd
,label=left:\text{\footnotesize $\begin{array}{c}\{\oldrk=\infty\}\end{array}$}
] at ($(0)+(0,-2)$) {$\ell_3$};
\node (4) [zustd
,label=left:\text{\footnotesize $\begin{array}{c}\{\oldrk=\infty\}\end{array}$}
] at ($(0)+(0,-4)$) {$\ell_4$};
\node (5) [zustd
,label=left:\text{\footnotesize $\begin{array}{c}\{\oldrk=\infty\}\end{array}$}
] at ($(0)+(-1,-1)$) {$\ell_5$};
\node (3') [zustd,accepting
,label=right:\text{\footnotesize $\begin{array}{c}\{i-j < \oldrk\\ \;\land\; \oldrk \geq 0\}\end{array}$}
] at ($(0)+(2,-2)$) {$\ell_3'$};
\node (4') [zustd
,label=right:\text{\footnotesize $\begin{array}{c}\{i-j = \oldrk\\ \;\land\; i-j > 0\}\end{array}$}
] at ($(0)+(2,-4)$) {$\ell_4'$};
\draw [trans] ($(0) + (125:0.7)$) to (0);
\draw [trans,bend left=20] (0) to node {$\stsccol{i>0}{outer}$} (2);
\draw [trans,bend left=20] (2) to node {$\stsccol{j:=1}{outer}$} (3);
\draw [trans,bend left=40] (3) to node {$\stsccol{j<i}{inner}$} (4);
\draw [trans,bend left=40] (4) to node {$\stsccol{j++}{inner}$} (3);
\draw [trans,bend left=20] (3) to node {$\stsccol{j>=i}{outer}$} (5);
\draw [trans,bend left=20] (5) to node {$\stsccol{i--}{outer}$} (0);
\draw [trans] (2) to node {$\stsccol{j:=1}{outer}$} (3');
\draw [trans,bend left=40] (3') to node {$\stsccol{j<i}{inner}$} (4');
\draw [trans,bend left=40] (4') to node {$\stsccol{j++}{inner}$} (3');
\end{tikzpicture}
}}

\begin{example}
The figure on the right depicts a certified module $(\Psort_1,f,\mI)$ where $f$ is the ranking function $f(i,j)=i-j$ and $\mI$ is the mapping of locations to predicates indicated by writing the predicate beneath the location.
\end{example}

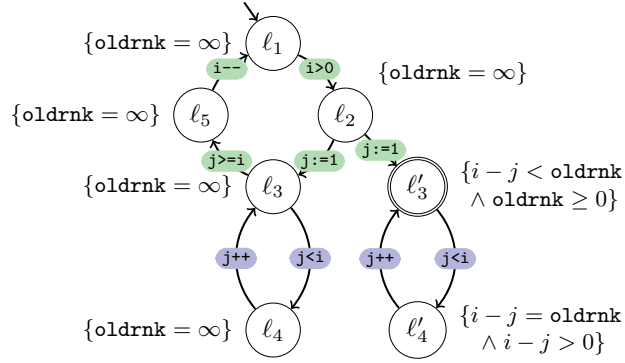
\begin{theorem}[soundness]
\label{thm:soundness}
Each fair $\omega$-trace of a certified module $(\prg,f,\mI)$ is terminating.
\end{theorem}

\section{Learning a terminating program}
\label{sec:automatic-module}

In this section we present a method for the construction of a certified module $(\prg, f, \mI)$.
The crux of this method is that we do not construct a termination argument (a ranking function $f$ together with a rank certificate $\mI$) for the resulting module $\prg$.
Instead, we construct vice versa the resulting module $\prg$ as the largest module for which a given termination argument (a ranking function $f$ together with a rank certificate $\mI$) is applicable.
We obtain this termination proof from a single  $\omega$-trace.
We call this method learning, because we learn a terminating program (given as a certified module) from a single sample trace.

The input to our method is a terminating $\omega$-trace $\st_1\dots\st_{k-1} ( \st_k\dots\st_{n} )^\omega $ that is ultimately periodic. 
We call an ultimately periodic trace a \defo{lasso}.
We call the prefix $\st_1\dots\st_{k-1}$ the \defo{stem} of the lasso and we call the periodic part $\st_k\dots\st_{n}$ the \defo{loop} of the lasso. 
For better legibility we use $u$ (resp. $v$) to denote the stem (resp. loop) of the lasso.
We construct a certified module $(\prg,f,\mI)$ in the following three steps. 

\subsection*{Step~1. Synthesize ranking function $f$}
First, we construct a module $\prg_{uv^\omega}$ that has only one single $\omega$-trace, namely the lasso $uv^\omega$. We call $\prg_{uv^\omega}$ the \defo{lasso module} of $uv^\omega$ and construct $\prg_{uv^\omega}=\langle \Loc,\delta, \ellinit, \{\ellfin\} \rangle$ formally as the module that has one location for each statement (i.e., $ \Loc=\{\ell_0,\dots,\ell_{n-1}\}$), where $\ell_0$ is the initial location, $\ell_k$ is the final location and the transition graph resembles the shape of a lasso, i.e., 
$\delta = \{ (\ell_i,\st_i,\ell_{i+1}) \mid i=1,\dots n-2\} \cup \{ (\ell_{n-1},\st_n,\ell_{k})\}$.

The lasso module $\prg_{uv^\omega}$ can be seen as a program that consists of a single while loop. 
This allows us to use existing methods~\cite{cav/Ben-Amram09,popl/Ben-AmramG13,cav/BradleyMS05,tacas/CookKRW10,conf/atva/HeizmannHLP13,conf/tacas/LeikeH14,vmcai/PodelskiR04} to synthesize a ranking function for $\prg_{uv^\omega}$.
% Furthermore, we can use any termination analysis which is able to generate ranking functions.

\parpic[r]{
\scalebox{1.0}{
\begin{tikzpicture}[scale=1,
zustd/.style={circle,draw,minimum size=10pt},
%trans/.style={>=stealth,thick}
trans/.style={thick,->}
]
\node (0) [zustd] at (0,0) {$\ell_1$};
\node (2) [zustd] at ($(0)+(2,0)$) {$\ell_2$};
\node (3) [zustd,accepting] at ($(0)+(4,0)$) {$\ell_3$};
\node (4) [zustd] at ($(0)+(6,0)$) {$\ell_4$};
\draw [trans] ($(0) + (125:0.7)$) to (0);
\draw [trans] (0) to node {$\stsccol{i>0}{outer}$} (2);
\draw [trans] (2) to node {$\stsccol{j:=1}{outer}$} (3);
\draw [trans,bend left=20] (3) to node {$\stsccol{j<i}{inner}$} (4);
\draw [trans,bend left=20] (4) to node {$\stsccol{j++}{inner}$} (3);
\end{tikzpicture}
}
}
\begin{example}
\label{ex:PsortConstr}
Given the $\omega$-trace $\qquad\qquad$
$\stsccol{i>0}{outer}\stsccol{j:=1}{outer} (\stsccol{j<i}{inner}\stsccol{j++}{inner})^\omega,$
we construct the lasso module $\prg_{uv^\omega}$ depicted on the right and synthesize the ranking function $f(i,j)=i-j$ for this module.
\end{example}
% \picskip{0}

\subsection*{Step~2. Compute rank certificate $\mI$}
\parpic[r]{
\begin{small}
$\begin{array}{c}
\hoare{true}{\stsc{$\oldrk$:=$\infty$}}{\mI(\ell_1)}\\
\hoare{\mI(\ell_i)}{\st_{i}}{\mI(\ell_{i+1})} \quad \text{ for } 1 \leq i < k\\[1mm]
\mI(\ell_{k}) \Rightarrow  f(\vec\var)<\oldrk\\[1mm]
\hoare{\mI(\ell_{k})}{\stsc{$\oldrk$:=$f(\vec\var)$}\;\st_{k}}{\mI(\ell_{k+1})}\\[1mm]
\hoare{\mI(\ell_i)}{\st_{i+1}}{\mI(\ell_{i+1})} \quad \text{ for } k < i < n\\[1mm]
\hoare{\mI(\ell_n)}{\st_n}{\mI(\ell_{k})}
\end{array}
$
\end{small}
}

Given the lasso module $\prg_{uv^\omega}$ and the ranking function $f$, we now compute a rank certificate $\mI$.
Since $\prg_{uv^\omega}$ has a ``lasso shape'' a mapping $\mI$ from the locations of $\prg_{uv^\omega}$ to predicates is a rank certificate if and only if the Hoare triples and the implication shown on the right are valid.

\parpic[l]{
\begin{minipage}{52mm}
\pseudo{program rankDecrease()}
$$\begin{array}{cl}
& \oldrk := \infty\\
\ell_1: & \st_1\\
% \ell_1: & \st_2\\
\vdots & \;\vdots\\
\ell_{k-1}: & \st_{k-1}\\
\ell_{k}: & \pseudo{while (true)} \\
& \quad\quad \pseudo{assert}(f(\vec\var)<\oldrk)\\
& \quad\quad \oldrk := f(\vec\var)\\
& \quad\quad \st_{k}\\
\ell_{k+1}: & \quad\qquad \st_{k+1}\\
\vdots & \quad\quad \;\vdots\\
\ell_{n}: & \quad\quad \st_{n}
\end{array}$$
\end{minipage}
}

The predicates  $\mI(\ell_i)$ for which these implications are valid, can be obtained by proving partial correctness of the program \rankDecrease\ depicted on the left.
The program \rankDecrease\ first assigns the value $\infty$ (which is strictly larger than any other element in the well-ordered set $\WF$) to the variable $\oldrk$.
Afterwards the statements $\st_1\dots\st_{k-1}$ are executed and the program \rankDecrease\ enters a nonterminating \pseudo{while} loop.
We use an assert statement to state the correctness specification of the program \rankDecrease.
The program is correct if at the beginning of the \pseudo{while} loop the inequality $f(\vec\var)<\oldrk$ holds.
After this assert statement, the current value of the function $f$ is assigned to the variable $\oldrk$ and then the statements $\st_{k}\dots\st_n$ are executed.

A Floyd-Hoare annotation $\mI(\ell_1),\dots,\mI(\ell_n)$ that shows partial correctness of the program \rankDecrease\ is also a rank certificate for our ranking function $f$ and our lasso module $\prg_{uv^\omega}$. 
This Floyd-Hoare annotation can be computed by static analysis~\cite{popl/CousotC77}.

\parpic[r]{
\scalebox{1.0}{
\begin{tikzpicture}[scale=1,
zustd/.style={circle,draw,minimum size=10pt},
%trans/.style={>=stealth,thick}
trans/.style={thick,->}
]
\node (0) [zustd
,label=below:\text{\footnotesize $\begin{array}{c}\{\oldrk=\infty\}\end{array}$}
] at (0,0) {$\ell_1$};
\node (2) [zustd
,label=above:\text{\footnotesize $\begin{array}{c}\{\oldrk=\infty\}\end{array}$}
] at ($(0)+(2,0)$) {$\ell_2$};
\node (3) [zustd,accepting
,label=below:\text{\footnotesize $\begin{array}{c}\{i-j \prec \oldrk\}\end{array}$}
] at ($(0)+(4,0)$) {$\ell_3$};
\node (4) [zustd
,label=above:\text{\footnotesize $\begin{array}{c}\{i-j \leq \oldrk\\ \;\land\; i-j\geq 0\}\end{array}$}
] at ($(0)+(6,0)$) {$\ell_4$};
\draw [trans] ($(0) + (125:0.7)$) to (0);
\draw [trans] (0) to node {$\stsccol{i>0}{outer}$} (2);
\draw [trans] (2) to node {$\stsccol{j:=1}{outer}$} (3);
\draw [trans,bend left=30] (3) to node {$\stsccol{j<i}{inner}$} (4);
\draw [trans,bend left=30] (4) to node {$\stsccol{j++}{inner}$} (3);
\end{tikzpicture}
}}

\begin{example}
Continuing Example~\ref{ex:PsortConstr} we construct the program \rankDecrease\ for $\prg_{uv^\omega}$ and compute the rank certificate depicted in the figure on the right. The rank certificate $\mI$ is represented by the predicates denoted beneath the locations.
\end{example}

\paragraph{An alternative variant of Step~2.}
Some methods for the synthesis of a ranking function \cite{cav/BradleyMS05,conf/atva/HeizmannHLP13} also provide a \emph{supporting invariant}. This is a predicate $I$ such that
\begin{itemize}
 \item $I$ is invariant under executions of the loop $\st_1\dots\st_{k-1}$,
 \item $I$ is an overapproximation of the reachable valuations after executing the stem $\st_k\dots\st_{n}$,
 \item and each execution of the loop starting in a valuation contained $I$ decreases the ranking function $f$.
\end{itemize}
If we have a supporting invariant $I$ for the ranking function $f$, we do not have to construct and analyze the program \rankDecrease.
% there is an alternative to the construction and the static analysis of the program \rankDecrease. 
Alternatively, we can set the predicate $\mI(\ell_k)$ to 
$$I\;\;\land\;\; f(\vec\var)<\oldrk\;\;\land\;\; \oldrk\geq 0$$
and obtain the remaining predicates $\mI(\ell_0),\dots, \mI(\ell_{k-1})$, and $\mI(\ell_{k+1}),\dots,\mI(\ell_n)$ as strongest postconditions by using an interpolating theorem prover.
%\cite{CHN12}.

\subsection*{Step~3. Construct module $\prg$} 
We extend the lasso module $\prg_{uv^\omega}$ to a module $\prg$ that also has the ranking function $f$ and that also has the rank certificate $\mI$.
Therefore we modify $\prg_{uv^\omega}$ according to the following two rules.

\begin{description}

 \item[Modification rule~1: merge locations]
 If the predicates mapped to the locations $\ell_i$ and $\ell_j$ coincide (i.e., $ \mI(\ell_i) = \mI(\ell_j) $) then we may merge both locations.
%  The result is still a certified module.
 \item[Modification rule~2: add transitions] Let $\st$ be some program statement and let $\ell_i$, and $\ell_j$ be locations.
 If $\ell_i \neq \ellfin$ and the Hoare triple $\hoare{\ell_i}{\st}{\ell_j}$ is valid, we may add the transition $(\ell_i,\st,\ell_j)$.
 If $\ell_i = \ellfin$ and the Hoare triple $\hoare{\ell_i}{\stsm{$\oldrk$:=$f(\vec\var)$};\st}{\ell_j}$ is valid, we may add the transition $(\ell_i,\st,\ell_j)$.
\end{description}
If we apply these modifications to a certified module we obtain again a certified module.
Every strategy for applying these modfications gives rise to an algorithm that is an instance of our method.
% If we apply Modifiation~2 until no more transitions can be added, we obtain the largest module $\prg$ for which the rank certificate $\mI$ proves that $f$ is a ranking function for $\prg$.

\parpic[r]{
\scalebox{1.0}{
\begin{tikzpicture}[scale=1,
zustd/.style={circle,draw,minimum size=10pt},
%trans/.style={>=stealth,thick}
trans/.style={thick,->}
]
\node (0) [zustd
,label=below:\text{\footnotesize $\begin{array}{c}\{\oldrk=\infty\}\end{array}$}
] at (0,0) {$\ell_{1}$};
\node (3) [zustd,accepting
,label=below:\text{\footnotesize $\begin{array}{c}\{i-j \prec \oldrk\}\end{array}$}
] at ($(0)+(3,0)$) {$\ell_3$};
\node (4) [zustd
,label=above:\text{\footnotesize $\begin{array}{c}\{i-j \leq \oldrk\\ \;\land\; i-j\geq 0\}\end{array}$}
] at ($(0)+(5,0)$) {$\ell_4$};
\draw [trans] ($(0) + (125:0.7)$) to (0);
\draw [trans,loop left] (0) to node {$\Sigma$} (0);
\draw [trans,auto] (0) to node {$\Sigma$} (3);
\draw [trans,bend left=30] (3) to node {$\stsccol{j<i}{inner}$} (4);
\draw [trans,bend left=30] (4) to node {$\stsccol{j++}{inner}$} (3);
\end{tikzpicture}
}}

\begin{example}
Continuing Example~\ref{ex:PsortConstr} we merge locations $\ell_1$ and $\ell_2$.
Afterwards we add for each program statement that occurs in $\Psort$ a selfloop at $\ell_1$ and a transition between $\ell_1$ and $\ell_3$.
We obtain the certified module $\prg_\mathsf{ext}$ depicted on the right.
The set of fair $\omega$-traces of this module is given by the $\omega$-regular expression $\Sigma^*.(\stsccol{j<i}{inner}\stsccol{j++}{inner})^\omega.$
If we take the intersection of the program $\Psort$ and the module $\prg_\mathsf{ext}$ 
% (if we \emph{refine} $\Psort$ by $\prg_\mathsf{ext}$) 
we obtain the module $\Psort_1$ from Example~1.  
In our algorithm (Section~\ref{sec:automatic-decomposition}), we do not need to construct modues such as $\Psort_1$ explicitly (we only use their implicit representation through $\prg_\mathsf{ext}$).
\end{example}

\section{Overall algorithm}
\label{sec:automatic-decomposition}

Until now, we have formalized (and automated) one part of our method, which is to construct a terminating module from a given sample trace.
We still need to formalize (and automate) how to check that a set of modules covers all behaviours of the program.
We will say that   the program $\prg$ has a decomposition into the modules $\prg_1,\ldots,\prg_n$ if the set of $\omega$-traces of the program $\prg$ is the union of the set of fair  $\omega$-traces of the modules $\prg_1,\ldots,\prg_n$.

We can automate the check that indeed all cases are covered by reducing it to the inclusion between Büchi automata.  Both a program and a module are special cases of Büchi automata (where the set of states is the set of program locations and the set of final states contains all program locations respectively the final location $\ellfin$ only).  By definition, the $\omega$-traces of the program $\prg$ are exactly the infinite words accepted by the Büchi automaton $\prg$ (and form the language $\mL(\prg)$ recognized by $\prg$), and
 the fair $\omega$-traces of the module $\prg_i$ are exactly the infinite words accepted by the Büchi automaton  $\prg_i$  (and  form the language $\mL(\prg_i)$  recognized by  $\prg_i$), for $i=1,\ldots,n$.   The inclusion
 $$\mL(\prg) \subseteq \mL(\prg_1)\cup\dots\cup \mL(\prg_n)$$
can be checked by a model checker such as~\cite{journals/tse/Holzmann97} or by a tool for manipulating Büchi automata such as~\cite{cav/TsaiTH13}.

We will use Büchi automata also in order to prove that decomposing a program into certified modules is in principle a complete method for termination analysis.

\begin{theorem}[completeness]\label{thm:completeness}
If a program $\prg$ is terminating then it can be decomposed into a finite set of certified modules, i.e.,
there are certified modules $$(\prg_1,f_1,\mI_1),\dots,(\prg_n,f_n,\mI_n)$$ such that the following equality holds.
$$\mL(\prg) = \mL(\prg_1)\cup\dots\cup \mL(\prg_n)$$
\end{theorem}

\paragraph{Overall algorithm.}
Having reduced the check that a set of modules is a decomposition of a program, we are ready to present our algorithm for termination analysis, depicted below.
The algorithm iteratively constructs certified modules $(\prg_i,f_i,\mI_i)$ until all $\omega$-traces of the program are known to be terminating or we encounter an $\omega$-trace for which we cannot find a termination argument.

% \begin{minipage}{50mm}
\begin{algorithm}
\SetKwInOut{Input}{input}\SetKwInOut{Output}{output}
\Input{program $\prg$}
\Output{certified modules $(\prg_1,f_1,\mI_1),\dots,(\prg_n,f_n,\mI_n)$ }
\For{$n=0,1,2,\dots$}{
\eIf{$\mL(\prg)\;\nsubseteq\;\mL(\prg_1)\cup\dots\cup\mL(\prg_{n-1})$}{
take $\omega$-trace $u.v^\omega$ that is counterexample to inclusion\;
construct lasso module $\prg_{uv^\omega}$\;
$f_{n}$ := synthesizeRankingFunction($\prg_{uv^\omega}$)\;
\If{$f_{n} =$ no ranking function found}{
\Return ``unable to decide termination of $\prg$''
% ``decomposition failed'' + $uv^\omega$
}
$\mI_{n}$ := computeRankCertificate($f_{n}$, $\prg_{uv^\omega}$)\;
$\prg_{n}$ := extendCertifiedModule($ \prg_{uv^\omega},f_{n},\mI_{n}$)\;
} {
\Return %$\prg$ is terminating, decomposition $(\prg_1,f_1,\mI_1),\dots,(\prg_n,f_n,\mI_n)$
 ``$\prg$ is terminating,'' \hspace{100mm} \strut \ \ \ \ \ \ \ \ \ \ \ ``found decomposition $(\prg_1,f_1,\mI_1),\dots,(\prg_n,f_n,\mI_n)$''
}
% take $u.v^\omega\in\mL(\prg)\backslash(\mL(\B_1)\cup\dots\cup\mL(\B_n))$\;
}
\caption{decomposition of a program $\prg$ into certified modules }
\label{algo:decomp}
\end{algorithm}
% \end{minipage}

At the beginning of each iteration (line 2) we check if there is an $\omega$-trace of the program $\prg$ that is not already a fair $\omega$-trace of one of the modules $\prg_1,\dots,\prg_{n-1}$ (for which termination has already been proven). % We note that this is an automata theoretic task which does not require any knowledge about the program semantics. This task can be solved by a model checker like ~\cite{journals/tse/Holzmann97}.
As mentioned above, we reduce this check to language inclusion of Büchi automata. Therefore we know that whenever there exists a counterexample to language inclusion there exists also a lasso-shaped counterexample. We take such a lasso-shaped $\omega$-trace $uv^\omega$ and construct a program (called lasso module) whose only $\omega$-trace is $uv^\omega$ (see Step~1 in Section~\ref{sec:automatic-module}). Next, we analyze termination of the lasso module $\prg_{uv^\omega}$. If we cannot find a ranking function $f_{n}$ for $\prg_{uv^\omega}$ our algorithm is unable to decide termination of $\prg$ and returns.
Otherwise we take a ranking function $f_{n}$ and construct a rank certificate $\mI_{n}$ for $f_{n}$ and $\prg_{uv^\omega}$ (see Step~2 in Section~\ref{sec:automatic-module}).
Afterwards we use the rank certificate to construct the module $\prg_{n}$. Termination of each fair $\omega$-trace of $\prg_{n}$ can be shown using the ranking function $f_{n}$ and the rank certificate $\mI_{n}$, i.e., $(\prg_{n},f_{n},\mI_{n})$ is a certified module (see Step~3 in Section~\ref{sec:automatic-module}).
If we were not able to find a counterexample to inclusion in line~2, the program $\prg$ is already decomposed into certified modules. We have proven termination and return the certified modules $(\prg_1,f_1,\mI_1),\dots,(\prg_n,f_n,\mI_n)$.

Our approach lends itself to a variation of the above algorithm where one uses an exit condition different from the inclusion check in line~2.  In that case, the algorithm returns the modules $\prg_1,\dots,\prg_{n-1} $constructed so far and, in addition, a ``remainder program'' $\prg_\mathsf{rem}$ which is constructed via the language-theoretic difference of Büchi automata.
$$\prg_\mathsf{rem} := \prg \backslash (\prg_1\cup\dots\cup \prg_{n-1})$$
This is interesting in a variety of contexts, e.g., when we found an $\omega$-trace that is nonterminating, or when we found an $\omega$-trace whose termination analysis failed, or simply in case of a timeout.  The remainder program can then be analyzed manually, or it can be used as a runtime monitor, etc.

\section{Evaluation}
\label{sec:evaluation}

It is unlikely that one approach outperforms all others on all kinds of programs, either in effectiveness (how many termination problems can be solved?) or in efficiency (... in what time?).
In this paper, we have presented the base algorithm of a new approach to termination analysis.
To explore optimizations and possibilities of integration with other approaches must remain a topic of future work.

The question is whether our approach is a potentially useful addition to the portfolio of existing approaches.
Therefore, the goal of the present experimental evaluation must be restricted to showing that the approach has a practical potential \emph{in principle}, regarding effectiveness and regarding efficiency.  
This is not obvious since there are at least two ``mission-critical'' questions, namely:
\begin{itemize}
 \item Will the algorithm just learn one terminating program $\prg_1, \prg_2, \ldots$ after the other, going through an infinite (or just unrealistically high) number of sample traces $\pi_1, \pi_2, \ldots$ ?
 \item Will the check of inclusion between Büchi automata (which is notoriously difficult and still an object of ongoing work~\cite{fossacs/BreuersLO12,conf/wia/TsaiFVT10}) be a `bad' bottleneck?
\end{itemize}

We put the evaluation into the context of a previous, very thorough evaluation\footnote{\url{http://verify.rwth-aachen.de/brockschmidt/Cooperating-T2/}} in~\cite{conf/cav/BrockschmidtCF13} that contained 260 terminating programs.
Out of the 260 programs, our tool can handle 236 programs. This, we believe, indicates the potential effectiveness of our approach.
In comparison regarding effectiveness, \textsc{Cooperating-T2}, the “winner” of the evaluation in~\cite{conf/cav/BrockschmidtCF13} (a highly optimized tool which integrates several approaches) can handle 14 programs that our tool cannot handle, but our tool can handle 5 programs that \textsc{Cooperating-T2} cannot handle (namely \textsf{\small a.10.c.t2.c}, \textsf{\small eric.t2.c}, \textsf{\small sas2.t2.c}, \textsf{\small spiral.t2.c} and \textsf{\small sumit.t2.c}). This confirms our point that no single approach provides a ``silver bullet'' and that it is desirable to have a large portfolio of approaches.

% Regarding efficiency, we decided against a pointwise comparison because the exact timings depend on a number factors and are therefore somewhat erratic (as can be seen from the demonstration category on termination at the SV-COMP\footnote{\url{http://sv-comp.sosy-lab.org/2014/results/}} ~\cite{conf/tacas/Beyer14}, its results, we think, should not be over-interpreted).  
% The overall timings of \textsc{Cooperating-T2} and our tool are in the same range 
% % (which is also the same range of most other tools in the evaluation in~\cite{...}.  
% In fact, the issue in termination analysis is not the scalability in terms of raw size of code but rather scalability in what one might call ``loop complexity''.  
% Loop complexity can obviously not be a metric based on syntax (such as, e.g., the nesting level of loops in the control flow graph). 
% To see this, take a single while loop with an if-then-else statement; its complexity depends on the number of possibilities to nest the then and else parts in  loop unfolding (which again depends on the evaluations of the the if-expression).   
% It is a topic of future work to design a metric for loop complexity.  
% One possible direction is to investigate the number of modules for this purpose.  
% % Indeed, the number of modules seems relatively high in the programs which \textsc{Cooperating-T2} cannot handle (such as ....). 

We implemented the algorithm presented in Section~\ref{sec:automatic-decomposition} in the tool \textsc{Ultimate BuchiAutomizer} that analyzes termination of C programs.
The input programs and the modules are represented by Büchi automata. In order to support (possibly recursive) functions, we use Büchi automata over nested words~\cite{journals/jacm/AlurM09} (we do not introduce the formalism in order to avoid the notational overhead) and implemented an automata library for these automata.
We do not check the inclusion 
\[\mL(\prg) \subseteq \mL(\prg_1)\cup\dots\cup \mL(\prg_n)\]
directly. Instead, we complement the modules and check the emptiness of their intersection with the program 
\[\mL(\prg) \cap \mL(\overline{\prg_1})\cup\dots\cup \mL(\overline{\prg_n})\]
which allow us to reuse intermediate results in further iterations. For complementing our Büchi automata we extended\cite{wu:11} the rank-based approach~\cite{conf/atva/FriedgutKV04} to (Büchi) nested word automata. The sample $\omega$-traces whose termination we analyze are obtained as counterexamples of an emptiness check that is implemented in our automata library. This emptiness check is purely automata theoric, does not exploit any information about the program, but prefers short counterexamples.
We use the tool \textsc{LassoRanker}~\cite{conf/atva/HeizmannHLP13,conf/tacas/LeikeH14} to synthesize ranking functions and supporting invariants for lassos. 
The Floyd-Hoare annotation is obtained via interpolation (alternative variant of Step~2 in Section~\ref{sec:automatic-module}).
For interprocedural $\omega$-traces we resort to \emph{nested interpolants}\cite{conf/popl/HeizmannHP10}. 
As interpolating theorem prover we use SMTInterpol~\cite{conf/spin/ChristHN12}. 
While constructing the modules, we apply Modification rule~1 (merge locations) always and we apply Modification rule~2 (add transitions) lazily in the following sense. Only if the automata library queries the existence of a transition in the module, we check whether this transition can be added by applying Modification rule~2.
Our tool is available as a command line version for download as well as via a web interface at the following URL.
\begin{center}
\url{http://ultimate.informatik.uni-freiburg.de/BuchiAutomizer/}
\end{center}

The following table shows the results for a subset of the benchmarks from~\cite{conf/cav/BrockschmidtCF13} where our tool run on a computer with an Intel Core i5-3340M CPU with 2.70GHz. Our tool and as well as \textsc{LassoRanker} the SMT solver, and the automata library are written in Java. The maximum heap size of the Java virtual machine was set to 4GB (\texttt{-Xmx4G}). 

For each example we list the lines of code of this example,
the overall runtime that our tool needed and the
time that our tool spend for analyzing lassos, constructing modules, and checking language inclusion of Büchi automata. 
Furthermore, we list the number of certified modules that had a trivial ranking function (e.g., $f(x)=0$), the number of certified modules that had a non-trivial ranking function, and the number of states of the largest module that was constructed.

\begin{center}
\strut

\strut

\strut

\strut

\begin{small}
\begin{tabular}{l@{\;\;\;}|@{\;\;\;}r@{\;\;\;}|@{\;\;\;}r@{\;\;\;}|@{\;\;\;}r@{\;\;\;}|@{\;\;\;}r@{\;\;\;}|@{\;\;\;}r@{\;\;\;}|@{\;\;\;}r@{\;\;\;}|@{\;\;\;}r@{\;\;\;}|@{\;\;\;}r}
filename & 
\begin{rotate}{90}\shortstack[1]{program\\ size}\end{rotate} &
\begin{rotate}{90}\shortstack[1]{overall\\ runtime}\end{rotate} &
\begin{rotate}{90}\shortstack[1]{lasso analysis\\ time}\end{rotate} &
\begin{rotate}{90}\shortstack[1]{module constr.\\ time}\end{rotate} &
\begin{rotate}{90}\shortstack[1]{Büchi inclusion\\ time}\end{rotate} &
\begin{rotate}{90}\shortstack[1]{modules\\ trivial rf}\end{rotate} &
\begin{rotate}{90}\shortstack[1]{modules\\ non-trivial rf}\end{rotate} &
\begin{rotate}{90}\shortstack[1]{module size\\ (maximum)}\end{rotate}
\\ \hline
a.10.c.t2.c 	     &  183 &   9s &  2.8s & 0.7s &  2.1s &   2 &  9 &  5\\
bf20.t2.c 	     &  156 &   6s &  0.7s & 0.9s &  1.9s &   6 &  7 &  9\\
bubbleSort.t2.c      &  109 &   5s &  0.7s & 0.3s &  1.2s &   5 &  5 &  5\\
consts1.t2.c 	     &   40 &   2s &  0.3s & 0.1s &  0.2s&    2 &  1 &  5\\
edn.t2.c 	     &  294 & 119s & 18.8s & 7.7s & 89.0s & 141 & 15 & 58\\
eric.t2.c 	     &   53 &  10s &  1.1s & 1.7s &  5.0s &   4 &  6 & 14\\
firewire.t2.c 	     &  178 &  28s &  3.6s & 1.3s & 19.0s &  12 &  7 &  8\\
mc91.t2.c 	     &   47 &  12s &  1.2s & 0.6s &  4.3s &   4 & 10 &  8\\
p-43-terminate.t2.c  &  727 & 124s &  2.1s & 4.2s & 110.6s &  6 & 18 &  5\\
reverse.t2.c 	     & 1351 &  14s &  3.1s & 1.2s &  2.9s &   2 &  3 & 12\\
s3-work.t2.c  	     & 3229 &  28s &  2.1s & 4.1s & 11.5s &   6 & 12 & 22\\
sas2.t2.c 	     &  192 &  12s &  1.3s & 3.0s &  5.5s &  12 &  6 & 17\\
spiral.c 	     &   65 &  38s &  0.9s & 1.3s & 32.7s &   8 & 12 & 14\\
sumit.t2.c 	     &   83 &   4s &  1.0s & 0.2s &  0.7s &   4 &  2 &  4\\
traverse\_twice.t2.c & 1428 &  12s &  1.7s & 1.4s &  3.2s &   2 &  4 & 18\\
ud.t2.c 	     &  279 &  32s &  2.1s & 3.8s & 22.1s &  30 & 25 & 32\\
\end{tabular}
% }
\end{small}
\end{center}

More results\footnote{\url{http://sv-comp.sosy-lab.org/2014/results/}} of our tool can be found at the SV-COMP 2014~\cite{conf/tacas/Beyer14} where our tool participated in the demonstration category on termination.

\paragraph{Discussion.} 
A reader who is familiar with Büchi automata may wonder why it is feasible to complement Büchi automata of these sizes. 
The answer lies in the flexibility that our definition of a module allows. 
We tuned the construction of modules in a way that the ``amount of nondeterminism'' is kept low.
However, it is still part of our future work to find a class of Büchi automata that can be easily complemented but does not hinder the module from accepting many traces. 
% This is also reflected by the number of different (non-trivial) ranking functions and the number of modules (with non-trivial ranking functions). Both numbers do not coincide. A single certified modules $(\prg,f,\mI)$ does not have to be a termination proof for \textit{all}  traces for which $f$ is a valid ranking function.

\section{Related work}
Our method is related to control flow refinement~\cite{pldi/GulwaniJK09}. There, a multi-path
loop is transformed into a semantically equivalent code fragment with simpler loops.
For example, following the algebraic decomposition rule %\ \mbox{$(a+b)^*=a^*ba^* + b^*$,} \
\[
(a+b)^*=(b^*ab^*)^+ + b^*
\]
the loop with the choice of two paths $a$ and $b$ is
transformed into the nondeterministic choice of two loops, one where
$a$ appears and one where it does not.

% The original incentive for our work was the control flow refinement~\cite{pldi/GulwaniJK09} that was introduced for cost analysis and safety.
% There, the control flow is refined according to algebraic rules. E.g., a loop
% $(a+b)^*$ is decomposed into modules
% $$(a+b)^* = \varepsilon + a^+ + b^+ + a^+b(a+b)^* + b^+a(a+b)^*.$$
% Then each module is checked independently.  
We extend control flow refinement by adding fairness
constraints~\cite{apal/Vardi91} and our reasoning is based on $\omega$-regular languages.  In
our running example (if we read $a$ as the outer and $b$ as the inner loop) we
decomposed the $\omega$-regular expression describing the nested loops
as follows %$(a+b)^\omega$ into
$$(a+b)^\omega = (a+b)^* b^\omega + (b^*a)^\omega.$$ 
We do not enforce the use of a fixed set of algebraic decomposition rules.  
Instead, we propose an algorithm that builds a decomposition on demand
from simple termination arguments.
Thus, we partition a set of traces only when it is necessary and,  \emph{by construction}, we
produce only modules that are guaranteed to have a simple termination argument.
%In our example, we could have swapped the inner and outer loop in the two termination arguments, yielding
%the decomposition
%$$(a+b)^\omega = (a+b)^* a^\omega + (a^*b)^\omega\,.$$
%However, in this decomposition the second loop $(a^*b)^\omega$ does
%not have a simple termination argument.

There are many other termination analyses, e.\,g., \cite{sefm/BabicHRC07,conf/tacas/CookSZ13,popl/CousotC12,cav/GantyG13,rta/GieslTSF04,conf/pldi/GrebenshchikovLPR12,conf/esop/UrbanM14}.  
Most related  are the termination analyses based on transition invariants and 
termination analyses based on size-change termination.

Termination analyses based on transition
invariants~\cite{conf/cav/BrockschmidtCF13,pldi/CookPR06,cacm/CookPR11,sas/HarrisLNR10,cav/KroeningSTW10,cav/LeeWY12,lics/PodelskiR04,popl/PodelskiR05}
combine different, independently obtained ranking functions to a termination argument. 
Using transition invariants it is sufficient to cover finite repetitions of the loop.
In our running example, one could cover the loop by
$$(a+b)^+ = b^+ + (b^*ab^*)^+$$
using the same simple ranking functions as our method for each case.  
Covering only finite traces is sound, as it can be shown that
$$(a+b)^\omega = (a+b)^* \; b^\omega + (a+b)^* \; (b^*ab^*)^\omega$$
using Ramsey's Theorem.
In our approach,  instead of having to
introduce $(a+b)^*$, we can get a more precise characterization
of the code before the infinite loop; also, we can  base our case-distinction
on which path was taken \emph{before} the loop was reached.
Furthermore, we 
get smaller expressions.  Compare the
expression $$(a+b)^*(b^*ab^*)^\omega$$ with our expression
$(b^*a)^\omega$.  Although they describe exactly the same traces, our
expression is simpler and therefore leads to a simpler termination
proof.
Redefining the loop entry point or unfolding
the loops are intrinsic techniques in our approach (as opposed to add-on heuristics).  If for the program $(ab)^\omega$, it is simpler
to prove the correctness of the loop $(baba)$, we  use
the fact that
$$(ab)^\omega = a(baba)^\omega.$$

The idea of size-change
termination~\cite{cav/Ben-Amram09,journals/corr/abs-1110-6183,popl/LeeJB01}
is to track the value of (auxiliary) variables and show the absence of infinite executions by showing that one value would be decreased infinitely often in a well-ordered domain.
The (auxiliary) variables can be seen as a predefined set of mutually
independent termination arguments. 
%The original presentation~\cite{popl/LeeJB01} used also Büchi automata to represent impossible program behaviors.

In contrast with the above approaches,  % cases in our setting are mutually
% independent, i.e., one can conclude the termination of the program
% solely from the fact that each case has a termination argument.  Each
a termination argument in our setting is a \emph{stand alone} module (its validity is checked
for the corresponding fair $\omega$-traces, independently from all other program traces).  In contrast, a component of a
lexicographic ranking function, a disjunct of a transition invariant,
or a size-change variable
makes sense only as part of a global termination argument (whose
validity has to be checked for the global program).

Finally, we use ``learning'' as a metaphor rather than as a technical term, in contrast with the work in~\cite{lee2012termination} which uses machine learning for termination analysis.

\section{Conclusion and Future Work}
We have presented a algorithm for termination analysis that transforms a program into a nondeterministic choice of programs. 
Our transformation is not guided by the syntactic structure of the program, but by its semantics. 
Instead of decomposing the program into modules and analyzing termination of the modules, we construct modules that we learned from sample traces and that are terminating by construction.

The general idea of such a transformation is the same as for \emph{trace refinement}~\cite{conf/esop/MauborgneR05}: move disjunction over abstract values to the disjunction over sets of traces.  
The formalization of the shared idea and the exploration of its theoretical and practical consequences for program analyses is a topic of future work.

\bibliographystyle{abbrv}
\bibliography{main}

\section{Proofs}
\begin{proof}[of Lemma~\ref{lem:rankingfunc}]
Let $\st_1,\st_2,\dots$ be a fair $\omega$-trace of $\prg$.  By definition,
the final location $\ellfin$ is visited infinitely often, i.\,e., there
is an infinite sequence $k_1<k_2<\dots$ such that after $\st_{k_i}$
the final location is visited.  Assume that the fair $\omega$-trace is not
terminating. Then, there exists an infinite sequence of valuations $\nu_0,\nu_1,\ldots$ such that for each $i\in\mathbb{N}$ the pair $(\nu_i,\nu_{i+1})$ is contained in the transition relation of the statement $\st_i$.
By the definition of the ranking function $f$, its value
decreases every time final location is visited, i.\,e.,
$$f(\nu_{k_1}) \succ f(\nu_{k_2}) \succ \dots.$$ 
This is not possible since $f$ maps into a well-ordered set.  Hence
and every fair $\omega$-trace of $\prg$ is
terminating.
\end{proof}

\begin{proof}[of Theorem~\ref{thm:soundness}]
First, we show that $f$ is a ranking function for the module $\prg$, afterwards this theorem is a direct consequence of Lemma~\ref{lem:rankingfunc}.

Consider a finite path of the module $\prg$
$$\ell_0\stackrel{\st_{1}}{\rightarrow}\dots\stackrel{\st_{k}}{\rightarrow}\ell_k\stackrel{\st_{k+1}}{\rightarrow}\cdots\stackrel{\st_n}{\rightarrow}\ell_n$$
that starts in the initial location and visits the final location in the $k$-th step and in the $n$-th step, i.\,e., 
$$\ell_0 = \ellinit \qquad\text{ and }\qquad \ell_k=\ell_n=\ellfin.$$
Let $\nu_0,\ldots,\nu_n$ be a sequence of valuations such that each pair of successive valuations $(\nu_i,\nu_{i+1})$ is in the transition relation of the statement $\st_i$, i.e.,
$$\nu_0\stackrel{\st_{1}}{\rightarrow}\dots\stackrel{\st_{k}}{\rightarrow}\nu_k\stackrel{\st_{k+1}}{\rightarrow}\cdots\stackrel{\st_n}{\rightarrow}\nu_n.$$

Next, we show that the strict inequality $f(\nu_n) < f(\nu_k)$ holds. Therefore, we extend the valuation $\nu_i$ with a value for the auxiliary variable $\oldrk$. We define this value $\valoldrk_i$ as follows.
\[\valoldrk_i := \begin{cases}
  \infty & \mbox{if }\forall j< i. \ell_j \neq \ellfin, \\[3mm]
  f(\nu_{j}) \text{\begin{small}$\begin{array}{c}\text{ where $j$ is greatest index }\\ \text{such that } j < i \text{ and } \ell_j = \ellfin\end{array}$\end{small}} & \mbox{otherwise }
\end{cases}\]
This extended valuation $(\nu_i\cup\{\oldrk\mapsto \valoldrk_i\})$ is denoted by $\bar\nu_i$.
Now, we show by induction that for all indices $i$ of our automaton run the extended valuation $\bar\nu_i$ is contained in the invariant $\mI(\ell_i)$, i.e.,
$$ \qquad \bar\nu_i \in \mI(\ell_i) \qquad \text{ for \ \ } i=0\ldots n.$$

Induction basis $i=0$. 
The extended valuation $\bar\nu_0$ is an element of $\mI(\ellinit)$, because the initial value of $\oldrk$ is $\infty$ 
and the predicate $\oldrk=\infty$ is equivalent to the invariant $\mI(\ellinit)$.

\goodbreak

Induction step $i\leadsto i+1$. \begin{itemize}
 \item Case 1: $i$ is index of an accepting state:
 
By the induction hypothesis the extended valuation $\bar\nu_i$ is contained in $\mI(\ell_i)$. 
According to the definition of a rank certificate the predicate $\mI(\ell_{i+1})$ is a superset of the predicate $post(\mI(\ell_i),\stsm{$\oldrk$:=$f(\vec\var)$};\st_i)$. 
Above we defined the value $\valoldrk_{i+1} := f(\nu_i)$.
Hence, the extended valuation $\bar\nu_{i+1}$ is contained in $\mI(\ell_{i+1})$.
 \item Case 2: $i$ is not index of an accepting state:

By the induction hypothesis the extended valuation $\bar\nu_i$ is contained in $\mI(\ell_i)$. 
According to the definition of a rank certificate the predicate $\mI(\ell_{i+1})$ is a superset of the predicate $post(\mI(\ell_i),\st_i)$.
Since the auxiliary variable $\oldrk$ does not appear in the program it is not modified by the statement $\st_i$. According to the definition above, the value $\valoldrk_{i+1}$ coincides with the value $\valoldrk_i$. Hence, the extended valuation $\bar\nu_{i+1}$ is contained in the predicate $\mI(\ell_{i+1})$.
\end{itemize}

Let $k_0<k_1<\ldots<k_m$ be the ascending chain of indices such that $k_0=k$, $k_m=n$ and $\ell_{k_j}=\ellfin$ for all $j=0,\ldots m$.
Since $\nu_{k_j}\in\mI(\ell_{k_j})$ the strict inequality $f(\nu_{k_j})<\valoldrk_{k_j}$ holds. As defined above, the value $\valoldrk_{k_j}$
is defined as $\nu_{k_{j-1}}$, hence the following sequence is a descending chain
$$f(\nu_{k_0})>f(\nu_{k_2})>\ldots>f(\nu_{k_m})$$
and especially $f(\nu_k)>f(\nu_n)$ holds and. Therefore $f$ is a ranking function for $\prg$. Using Lemma~\ref{lem:rankingfunc}, we conclude that each fair $\omega$-trace of $\prg$ is terminating.
\end{proof}

\begin{proof}[of Theorem~\ref{thm:completeness}]
The proof procedes in two steps. First we show that a program can be decomposed into modules.  In the second step we
show that we can give a ranking function and rank certificate for each terminating module.

The theorem of Büchi says that we can decompose each $\omega$-regular language $L$ into a finite disjuction
$$L=\bigcup_{i=1}^n U_i.V_i^\omega$$
where each $U_i$ and each $V_i$ is a regular language.

Let $\mathcal{A}^U_i$ and $\mathcal{A}^V_i$ be finite deterministic automata that recognize the regular languages $U_i$ and $V_i$, respectively. 
We construct the module $\mathcal{P}_i$ using the standard construction where $\mathcal{A}^U_i$ and $\mathcal{A}^V_i$ are combined to a Büchi automaton that recognizes the language $U_i.V_i^\omega$. 
The (single) final state of $\mathcal{P}_i$ is the initial state of $\mathcal{A}^V_i$ in this construction.
Hence, we can decompose the program $\prg$ into fair modules
$$\mL(\prg) = \mL(\prg_1)\cup\dots\cup \mL(\prg_n).$$

Since the modules contain the same executions as the program $\prg$ they must also be terminating.
Existence of a computable ranking functions $f$ is a classical result.%~\cite{popl/CousotC12}.
We take such a ranking function and extend the module by a specification that asserts that this ranking function is decreasing whenever the final location is visited.
A Floyd-Hoare annotation which shows partial correctness of the extended module can be seen as rank certificate $\mI$. 
The existence of this Floyd-Hoare annotation is also a classical result~\cite{ABO09-Book3}.
\end{proof}

% \section{Discussion of Construction Step 2.}
% 
% The assert statement in program \rankDecrease\  is valid if and only if the function $f$ is a ranking function. This allows us to propose an alternative in Step~1. Instead of synthesizing a ranking function $f$, we can guess ranking function candidates. A wrong guess will be detected in Step~2 and we can continue by guessing a ranking function candidate using a different method. 
% 
% \medskip
% 
% 
% 
% \smallskip
% 
% In \cite{pldi/CookPR06} the authors also state validity of a termination argument by annotating the program with additional assertions and auxilliary variables.
% In their contruction there is one auxilliary variable for each program variable.
% Several ranking functions and nondeterminsitic assignments of auxilliary variables are used.
% The assert statement holds if for any number of loop executions some ranking function is decreasing.
% In our construction the assert statement is valid if the value of the ranking function decreases in each execution of the loop.

% \input{OldStuffNotUsedInSubmission}

\end{document}